\documentclass[prx,twocolumn,showpacs,longbibliography]{revtex4-1}
\usepackage[latin1]{inputenc}
\usepackage{amsmath,amssymb,graphicx}
\usepackage{color}
\usepackage{xspace}
\usepackage{graphicx}
\usepackage{amssymb}
\usepackage{amsmath}
\usepackage{xspace}
\usepackage{dcolumn}
\usepackage{bm}
\usepackage{color}
\usepackage[extra]{tipa}

\newcommand{\oh}{\mbox{$\frac{1}{2}$}}

\newcommand{\threeh}{\mbox{$\frac{3}{2}$}}

\newcommand{\las}[0]{\langle}
\newcommand{\ras}[0]{\rangle}
\newcommand{\llas}[0]{\langle\langle}
\newcommand{\rras}[0]{\rangle\rangle}
\newcommand{\la}[0]{\left\las}
\newcommand{\ra}[0]{\right\ras}
\newcommand{\ket}[1]{\left|#1\ra}
\newcommand{\bra}[1]{\la#1\right|}

\renewcommand{\tilde}[1]{\widetilde{#1}}

\newcommand{\ie}[0]{i.e.\@\xspace}
\newcommand{\eg}[0]{e.g.\@\xspace}

\newcommand{\rmi}{\text{i}}
\newcommand{\rmd}{\text{d}}
\newcommand{\UP}[0]{\uparrow}
\newcommand{\DO}[0]{\downarrow}

\newcommand{\om}[0]{\omega}

\newcommand{\kB}{k_\text{B}}
\newcommand{\nag}{{\phantom{\dag}}}

\newfont{\tensy}{cmsy10}
\newcommand{\chem}[1]{{$\fontdimen16\tensy=3.0pt
    \fontdimen17\tensy=3.0pt \mathrm{#1}$}}

\definecolor{col4}{rgb}{0.0,0.0,0.8}
\definecolor{col1}{rgb}{0.12,0.56,1}

\begin{document}

\title{Topological Invariant and Quantum Spin Models from Magnetic $\pi$ Fluxes\\ in
  Correlated Topological Insulators}

\author{F. F. Assaad, M. Bercx, and M. Hohenadler}

\affiliation{Institut f\"ur Theoretische Physik und Astrophysik,
    Universit\"at W\"urzburg, Am Hubland, 97074 W\"urzburg, Germany}

\begin{abstract} The adiabatic insertion of a $\pi$ flux into a quantum
    spin Hall insulator gives rise to localized spin and charge fluxon
    states. We demonstrate that $\pi$ fluxes can be used in exact quantum Monte Carlo
    simulations to identify a correlated $Z_2$ topological insulator using
    the example of the Kane-Mele-Hubbard model. In the presence of repulsive
    interactions, a $\pi$ flux gives rise to a Kramers doublet of spin fluxon states
    with a Curie law signature in the magnetic susceptibility. Electronic
    correlations also provide a bosonic mode of magnetic excitons with tunable
    energy that act as exchange particles and mediate a dynamical
    interaction of adjustable range and strength between spin fluxons. $\pi$ fluxes can
    therefore be used to build models of interacting spins. This idea is
    applied to a three-spin ring and to one-dimensional spin chains.
    Because of the freedom to create almost arbitrary spin lattices,
    correlated topological insulators with $\pi$ fluxes represent a novel kind of
    quantum simulator potentially useful for numerical simulations and experiments.
\end{abstract}


\date{\today}
\maketitle

\section{Introduction}

A topological insulator represents a novel state of matter characterized by a
special band structure that can result, \eg, from strong spin-orbit
interaction \cite{Mo10,HaKa10}. In two dimensions, this state is
called a {\it quantum spin Hall insulator}, and has deep connections
with the quantum Hall effect, including the coexistence of a bulk band gap
and metallic edge states, the absence of symmetry breaking, and the
possibility of a mathematical
classification \cite{KaMe05a,PhysRevB.78.195125}. Importantly, because
of the absence of a magnetic field, the quantum spin Hall insulator preserves
time-reversal symmetry which provides protection against interactions and
disorder \cite{KaMe05b,Wu06,Cenke06}.  The quantum spin Hall insulator has been realized in
HgTe quantum wells \cite{BeHuZh06,Koenig07}.

Correlated topological insulators with strong electron-electron interactions
are the focus of current research \cite{HoAsreview13}. Intriguing concepts
include electron fractionalization in the presence of time-reversal
symmetry \cite{PhysRevLett.98.186809,PhysRevLett.101.086801,PhysRevLett.108.046401,Fi.Ch.Hu.Ka.Lu.Ru.Zy.11}, spin
liquids \cite{Meng10,Hohenadler10,Fi.Ch.Hu.Ka.Lu.Ru.Zy.11}, and topological
Mott insulators \cite{RaQiHo08,PeBa10}. Remarkably, some of the theoretical models can be studied using exact numerical methods.
A central problem in this context is the question of how to detect a
topological state directly from bulk properties, for example in cases
where the bulk-boundary correspondence breaks down. Experimentally, this
issue also arises in the absence of sharp
edges in proposed cold atom realizations as a result of the trapping
potential \cite{PhysRevLett.107.145301,Sun.2012}. The classification in terms
of a $Z_2$ Chern-Simons index relies on Bloch wavefunctions and is therefore only valid for
noninteracting systems. Generalizations involve twisted boundary conditions \cite{PhysRevB.31.3372}
or Green functions \cite{PhysRevLett.105.256803,PhysRevB.83.085426,arXiv:1201.6431v2,arXiv:1203.1028,arXiv:1111.6250,Budichetal},
and are challenging to use in experiments or exact simulations.  Indirect
signatures such as the closing of gaps \cite{Hohenadler10} or the crossing of
energy levels \cite{Va.Su.Ri.Ga.11} require, among other difficulties, experimental
tuning of microscopic parameters. 

Topological insulators show a unique response to topological defects such
as dislocations \cite{Ran-dislocations,Zaanen12} or $\pi$
fluxes \cite{PhysRevLett.101.086801,Qi08,Zaanen12}.  Upon adiabatic insertion of
a $\pi$ flux, Faraday's law together with the quantized transverse
conductivity gives rise to midgap {\it charge and spin fluxon states} \cite{PhysRevLett.101.086801,Qi08}.
These states are exponentially localized around the flux \cite{PhysRevLett.101.086801,Qi08}.
The existence of these states is ensured even in the presence of interactions
or disorder by time-reversal symmetry, and has been suggested as a bulk probe of the $Z_2$
index \cite{PhysRevLett.101.086801,Qi08}. The concept of fluxons can
also be generalized to situations where spin is not conserved, such as in
the presence of Rashba coupling. In three dimensions, a magnetic flux gives
rise to the {\it wormhole effect} \cite{PhysRevB.82.041104}. Electron-electron
repulsion lifts the degeneracy of charge and spin fluxons, but the two
degenerate spin fluxon states constitute a localized spin
with $S^z=\pm1/2$ \cite{PhysRevLett.101.086801}. Dynamical $\pi$ fluxes emerge
in the context of fractionalized topological insulators \cite{PhysRevLett.101.086801,PhysRevLett.108.046401}.

Previous work on $\pi$ fluxes in {\it noninteracting} quantum spin Hall
insulators \cite{PhysRevLett.101.086801,Qi08,Zaanen12} was based on
square-lattice models such as that for HgTe quantum wells \cite{BeHuZh06}.
Here we consider the half-filled Kane-Mele model on the honeycomb lattice \cite{KaMe05a},
historically the first model with a $Z_2$ topological phase, which has
close connections to graphene \cite{KaMe05a}, the integer quantum Hall
effect \cite{Ha88}, and when including interactions, to correlated Dirac
fermions \cite{Meng10}. Topological phases of interacting fermions on
honeycomb lattices may  be realized in transition metal
oxides \cite{Irridates-Nagaosa}, semiconductor structures \cite{Singha11},
graphene \cite{PhysRevX.1.021001}, or cold atoms \cite{honeycomb-esslinger},
see also Ref.~\cite{HoAsreview13}.

Here we use $\pi$ fluxes in combination with exact quantum Monte Carlo
simulations, and show that they can be used efficiently to probe the
topological invariant of correlated topological insulators. In particular, this method does
not rely on an adiabatic connection to a noninteracting state, and
may also be used for fractional states.  In addition, we demonstrate that
$\pi$ fluxes permit the construction of quantum spin models of almost arbitrary
geometry and with tunable, dynamical interactions. The spins correspond to
the spin fluxons created by inserting $\pi$ fluxes, and the interaction is
mediated by magnetic excitons corresponding to collective magnetic
fluctuations of the topological insulator. These spin models can be studied
theoretically with the quantum Monte Carlo method, or experimentally. As
examples, we show that a ring of three spins has a ground state with
magnetization $1/2$, and that a one-dimensional chain of fluxons undergoes a
Mott transition and is described at low energies by an XXZ model.

The article is organized as follows. In Sec.~\ref{sec:model}, we introduce
the Kane-Mele and Kane-Mele-Hubbard models. Section~\ref{sec:method} provides
details about the methods. The use of $\pi$ fluxes as a probe for topological
states is discussed in Sec.~\ref{sec:probe}, whereas the construction of
quantum spin models is the topic of Sec.~\ref{sec:spinmodels}. Conclusions
are given in Sec.~\ref{sec:conclusions}, and we provide three appendices.

\begin{figure}[t]
\includegraphics[width=0.45\textwidth,clip]{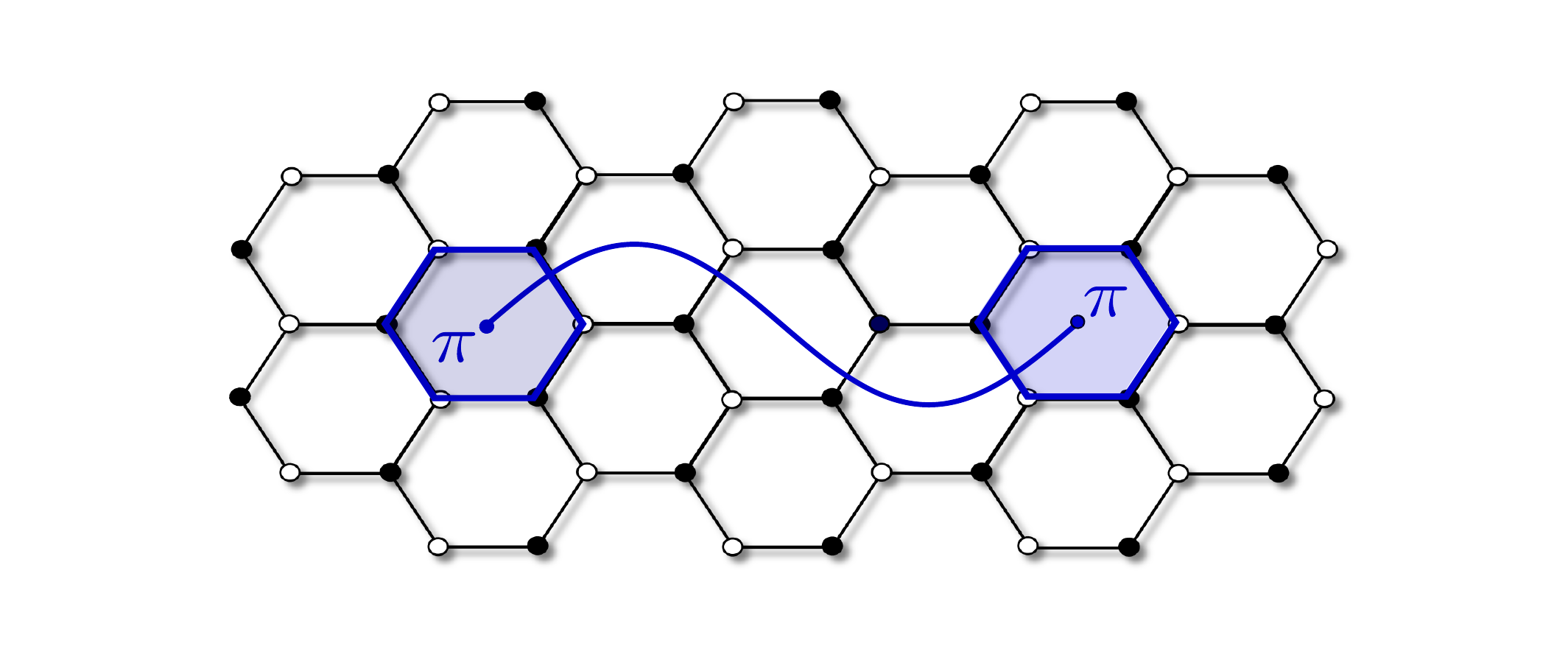}
\caption{\label{fig:branchcut}
(Color online)
For a lattice with periodic boundaries, $\pi$ fluxes
can be inserted in pairs. Each flux threads a hexagon (highlighted in
blue) of the honeycomb lattice, and the pair is connected by a branch cut
(blue line). Hopping processes crossing the branch cut acquire a phase
$e^{\rmi\pi}=-1$.
}
\end{figure}

\section{Model}\label{sec:model}

The half-filled Kane-Mele model with additional electron-electron
interactions can be studied with powerful quantum Monte Carlo
methods \cite{Hohenadler10,Zh.Wu.Zh.11}. Using the
spinor notation  ${\hat{c}^{\dagger}_{i} =
  \big(\hat{c}^{\dagger}_{i\uparrow},
  \hat{c}^{\dagger}_{i\downarrow}\big)}$, where
$\hat{c}^{\dagger}_{i\sigma}$ is a creation operator for an electron in a
Wannier state at site $i$ with spin $\sigma$, the Hamiltonian reads
\begin{align}\label{eq:KM}
  H_{\mbox{\scriptsize{KM}}} 
  = 
  -t \sum_{\las i,j \ras} \tau_{ij}    \hat{c}^{\dagger}_{i} \hat{c}^\nag_{j} 
  &+ 
  \rmi\,\lambda \sum_{\llas i,j\rras} \tau_{ij} \,
   \hat{c}^{\dagger}_{i}\,
   (\boldsymbol{\nu}_{ij} \cdot \boldsymbol{\sigma})   \,
   \hat{c}^\nag_{j} 
  \\\nonumber
  &\,
  +
  \rmi\,\alpha \sum_{\las i,j\ras}  \tau_{ij} \,
  \hat{c}^{\dagger}_{i}
  (\bm{s} \times \hat{\bm{d}}^\nag_{ij})\cdot \hat{\bm{z}} \,
  \hat{c}^\nag_{j}\,. 
\end{align}
The notation $\las i,j\ras$  and $\llas i,j\rras$ indicates that the sites $i$
and $j$ are nearest neighbors  and next-nearest neighbors, respectively,
and implicitly includes the Hermitian conjugate terms.  

The first term describes the hopping of electrons between neighboring
lattice sites. The second term represents the
spin-orbit coupling which reduces the $SU(2)$ spin rotation symmetry to a
$U(1)$ symmetry. The third term is an additional Rashba spin-orbit
coupling \cite{Rasha}. The additional
factors $\tau_{ij} = \pm 1 $ take into account any $\pi$ fluxes present,
whereas the original Kane-Mele model (without $\pi$ fluxes)
is recovered from Eq.~(\ref{eq:KM}) by setting $\tau_{ij}=1$.

The spin-orbit term corresponds to a 
next-nearest neighbor hopping with a complex amplitude $\rmi\lambda$,
and has been derived from the spin-orbit coupling in graphene \cite{KaMe05a}.
This hopping acquires a sign $\pm1$ depending on its direction,
the sublattice, and the electron spin. This sign is encoded in
$(\boldsymbol{\nu}_{ij} \cdot \boldsymbol{\sigma})$, where
\begin{equation}
  \boldsymbol{\nu}_{ij} = \frac{\bm{d}_{ik} \times \bm{d}_{kj}}{|\bm{d}_{ik} \times \bm{d}_{kj}|}\,,
\end{equation}
$\bm{d}_{ik}$ is the vector connecting sites $i$ and $k$, and $k$ is
the intermediate lattice site involved in the hopping process from $i$ to $j$.
For a coordinate independent representation, the vectors
$\bm{d}_{\alpha\beta}$ are defined in three dimensional space, although the $z$ component vanishes. The vector
$\boldsymbol{\sigma}$ is defined by $\boldsymbol{\sigma} = (\sigma^x,\sigma^y,\sigma^z)$,
with the Pauli matrices $\sigma^\alpha$.

The last term in Eq.~(\ref{eq:KM}) is the Rashba spin-orbit
interaction \cite{KaMe05a,KaMe05b}. It is defined in terms of the spin vector
$\bm{s}=\boldsymbol{\sigma}/2$, and the unit vector $\hat{\bm{d}}_{ij}$ which
can be expressed in terms of the nearest-neighbor vectors $\bm{\delta}_1$,
$\bm{\delta}_2$, $\bm{\delta}_3$ \cite{Neto_rev}.  The Rashba  coupling breaks the $z\mapsto-z$ inversion
symmetry, and has to be taken into account, for example, in the presence
of a substrate. Because this term includes spin-flip terms, spin
is no longer conserved. 
The Rashba term has been included in the results for the noninteracting
model~(\ref{eq:KM}), but cannot be included in quantum Monte Carlo simulations of
the interacting model~(\ref{eq:KMH}) due to a minus-sign problem.

The model~(\ref{eq:KM}) can be solved exactly
\cite{KaMe05a,KaMe05b,PhysRevLett.97.036808}. In the absence of Rashba
coupling, $\alpha=0$, the Kane-Mele model describes a $Z_2$ quantum spin Hall
insulator for any $\lambda>0$. This state is characterized by a bulk band gap
$\Delta_\text{sp}=3\sqrt{3}\lambda$, a spin gap
$\Delta_\text{s}=2\Delta_\text{sp}$, and a quantized spin Hall conductivity
$\sigma_{xy}^s=\frac{e^2}{2\pi}$. The topological state survives for Rashba
interactions $\alpha<2\sqrt{3}\lambda$ (for chemical potential $\mu=0$), and
has protected, helical edge states for geometries with open boundaries
\cite{KaMe05a,KaMe05b,PhysRevLett.97.036808}.
We use $t$ as the unit of energy ($\hbar=1$), take $\lambda/t=0.2$, and
consider periodic lattices with $L\times L'$ unit cells.

To investigate the effect of electron-electron repulsion,
we consider the paradigmatic Hubbard interaction \cite{Hu63} and arrive at the
{\it Kane-Mele-Hubbard model} \cite{RaHu10},
\begin{align}\label{eq:KMH}
  H_{\mbox{\scriptsize{KMH}}} = H_\text{KM} +  H_U\,,\quad
  H_U = \oh U
  \sum_{i} (\hat{c}^{\dagger}_{i} \hat{c}^\nag_{i} -  1 )^2\,.
\end{align}
Hamiltonian~(\ref{eq:KMH}) {\it without} Rashba coupling has been studied
intensely \cite{RaHu10,Hohenadler10,Zh.Wu.Zh.11,Gr.Xu.11,Yu.Xie.Li.11,Wu.Ra.Li.LH.11,PhysRevLett.107.166806,Ho.Me.La.We.Mu.As.12}.
In particular, its symmetries permit the application of exact quantum Monte Carlo methods without a sign
problem \cite{Hohenadler10,Zh.Wu.Zh.11,Ho.Me.La.We.Mu.As.12}.

On a lattice with periodic boundaries, $\pi$ fluxes can only be
inserted in pairs, as illustrated for the minimal number of two fluxes in
Fig.~\ref{fig:branchcut}. The flux pair is connected by a branch cut (or string),
and every hopping process crossing the cut acquires a phase $e^{\rmi\pi}=-1$, as encoded by 
$\tau_{ij}$ in Eq.~(\ref{eq:KM}). Different choices of the branch
cut for fixed flux positions are related by a gauge transformation.

\section{Method}\label{sec:method}

We have used the auxiliary-field quantum Monte Carlo method \cite{Hirsch81},
which was previously applied to the Hubbard model on the honeycomb
lattice \cite{Meng10}, and the Kane-Mele-Hubbard model \cite{Hohenadler10,Zh.Wu.Zh.11,Ho.Me.La.We.Mu.As.12}.
The central idea of this stochastic method is to use a path integral
representation of the interacting model~(\ref{eq:KMH}). By means of a
Hubbard-Stratonovich transformation, the Hubbard term is decoupled, leading
to a problem of noninteracting fermions in an external, space and
imaginary-time dependent field. The sampling is over different configurations
of these auxiliary fields in terms of local updates. For a given
configuration of fields, Wick's theorem can be used to calculate arbitrary
correlation functions from the single-particle Green function. We refer to
a review \cite{Assaad08_rev}, and previous work
\cite{Meng10,Hohenadler10,Ho.Me.La.We.Mu.As.12} for technical details
such as the calculation of energy gaps.

Here we have used a projective formulation (with projection parameter
$\theta t=40$) to obtain ground state results starting from a trial
wavefunction (the ground state of the $U=0$
case) \cite{Ho.Me.La.We.Mu.As.12}, and a finite-temperature formulation to
calculate thermodynamic properties. Both variants rely on a Trotter
discretisation of imaginary time (we used $\Delta\tau=\beta/L=\theta/L=0.1$), but the associated
systematic error is smaller than the statistical errors. At half filling,
time-reversal invariance ensures that simulations can be carried out without
a minus-sign problem even in the presence of $\pi$ fluxes.

\section{Using $\pi$ fluxes to probe correlated topological states}\label{sec:probe}

\begin{figure}[t]
\includegraphics[width=0.45\textwidth,clip]{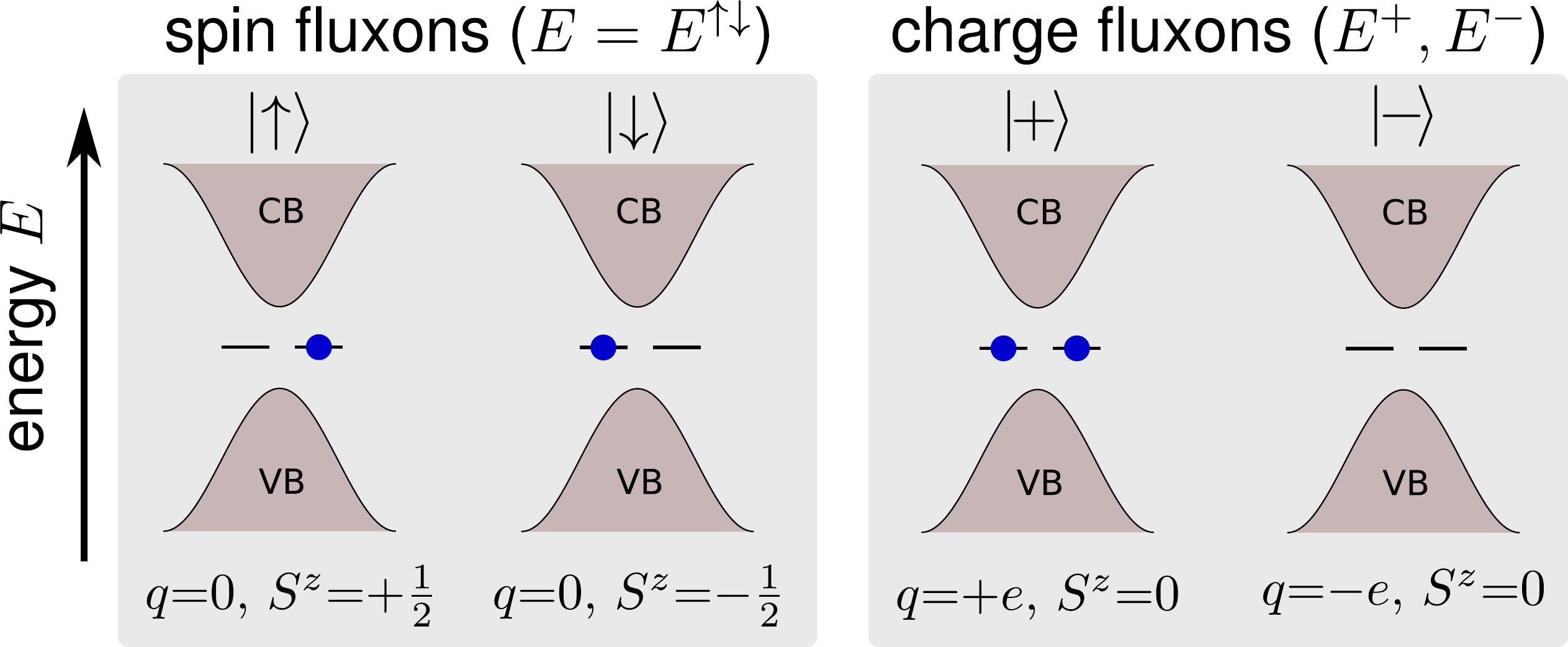}
\caption{\label{fig:fluxes}
(Color online)
In a quantum spin Hall insulator, a $\pi$ flux gives rise to four
states (with charge $q$ and spin $S^z$) localized near the flux, 
which lie inside the bulk energy gap between the valence and conduction
bands (labeled ``VB'' and ``CB'' in the figure, respectively) \cite{PhysRevLett.101.086801,Qi08}. The states
correspond to a Kramers doublet of spin fluxons
$\ket{\UP},\ket{\DO}$ with energy $E^{\UP\DO}$, and a doublet of charge
fluxons $\ket{+},\ket{-}$ with energies $E^+$, $E^-$. 
}
\end{figure}

\subsection{Thermodynamic signature of $\pi$ fluxes}

In the topological phase of the model~(\ref{eq:KM}), each $\pi$ flux
gives rise to four fluxon states which are exponentially localized (due to
the bulk energy gap $\Delta_\text{sp}$) near the corresponding flux-threaded
hexagons \cite{PhysRevLett.101.086801,Qi08}, see Fig.~\ref{fig:branchcut}.  The states correspond to the spin fluxons
$\ket{\UP},\ket{\DO}$ with energy $E^{\UP\DO}$, forming a Kramers pair 
related by time reversal, and the charge fluxons
$\ket{+},\ket{-}$ (with energies $E^+,E^-$) related by particle-hole
symmetry. As we show in Fig.~\ref{fig:fluxes}, the fluxon states lie inside
the bulk band gap, and for noninteracting electrons $E^{\UP\DO}=E^+=E^-$. 

The fluxons leave a characteristic signature in the static spin and charge
susceptibilities
\begin{equation}\label{eq:susc}
  \chi_\text{s}  
  = 
  \beta \left( \langle \hat{M}_z^2 \rangle - \langle \hat{M}_z \rangle^2 \right)
  \,,\quad   
  \chi_\text{c}  
  = 
  \beta \left( \langle \hat{N}^2 \rangle - \langle \hat{N} \rangle^2  \right)
  \,,
\end{equation}
which are defined in terms of the 
operators of total spin in the $z$ direction, $ \hat{M}_z = \sum_{i}
\hat{c}^{\dagger}_{i} \sigma^{z} \hat{c}^\nag_{i} $, and of the total
charge, $\hat{N}=\sum_{i}
\hat{c}^{\dagger}_{i} \hat{c}^\nag_{i} $; the inverse temperature is given by $\beta=\frac{1}{\kB T}$.
At low temperatures $\kB T \ll \Delta_\text{sp}$, we can restrict the
Hilbert space to $\{\ket{\UP},\ket{\DO},\ket{+},\ket{-}\}$. If the spin fluxons are
independent, and for $\alpha=0$, we expect a Curie law  $\chi_\text{s} = \chi_\text{c}
=\frac{1}{2 \kB T}$ per $\pi$ flux, and hence $\chi_\text{s} = \chi_\text{c}
= \frac{1}{\kB T}$ for two independent $\pi$ fluxes (see Appendix~\ref{app:twospins}). The
prefactor of the Curie law follows from the quantized spin Hall conductance
in the absence of Rashba coupling \cite{KaMe05a}.  Similarly, a Curie law 
was also predicted for topological excitations in polyacetylene \cite{PhysRevLett.42.1698}.

\begin{figure}[t]
\includegraphics[width=0.45\textwidth]{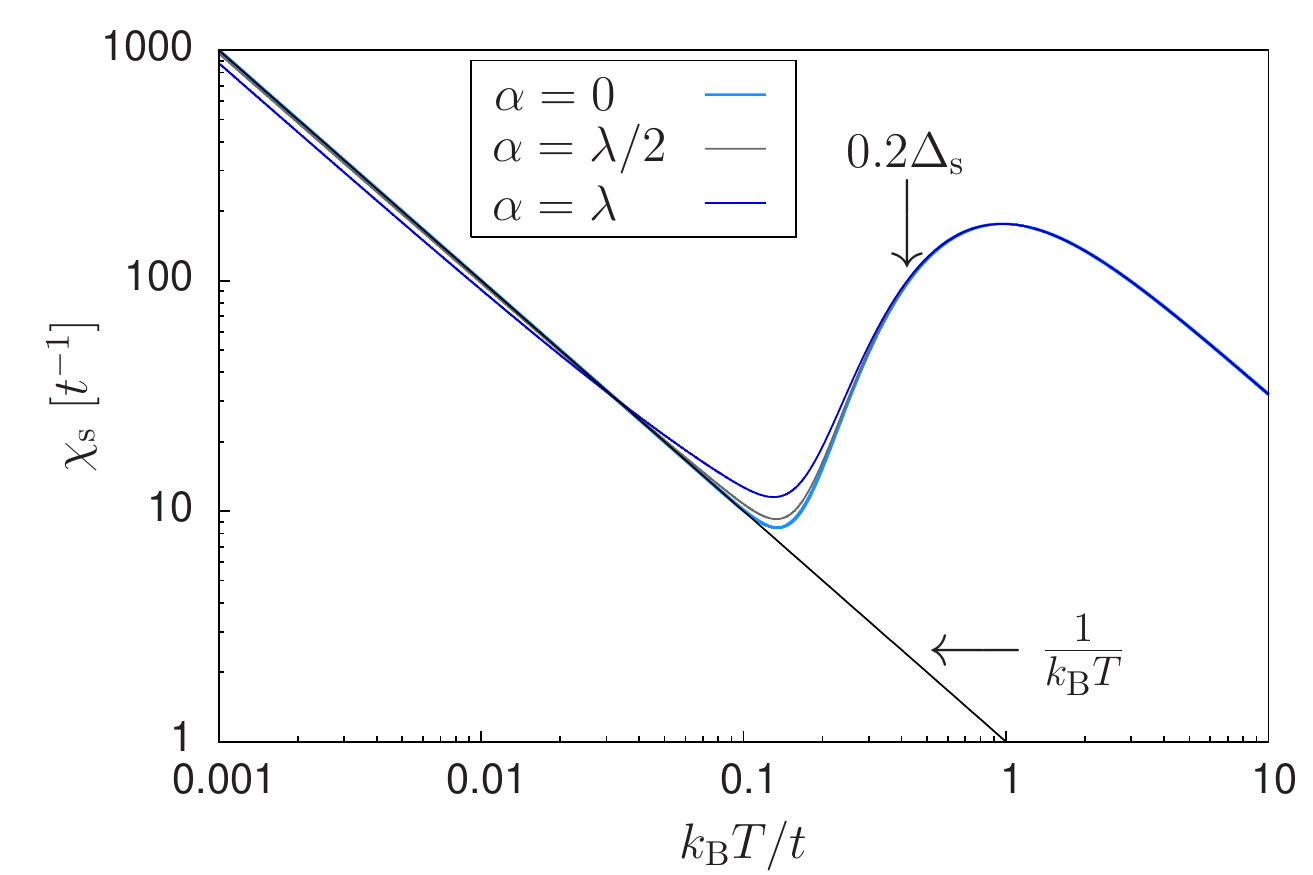}
\caption{\label{fig:freespinons}
(Color online)
Spin susceptibility of the Kane-Mele
model ($\lambda/t=0.2$) with two $\pi$ fluxes at the maximal distance on an
$18\times18$ lattice, for different Rashba couplings $\alpha$. At temperatures
$\kB T\lesssim 0.1 t$, each $\pi$ flux contributes $\frac{1}{2\kB T}$ to the
susceptibility, leading to $\chi_\text{s}\approx\frac{1}{\kB T}$.  Also shown
is the spin gap energy scale $0.2\Delta_\text{s}$ for $T=0$, $\alpha=0$. For
$\alpha>0$, the chemical potential is adjusted to retain a half-filled band.
}
\end{figure}

\begin{figure}
\includegraphics[width=0.45\textwidth]{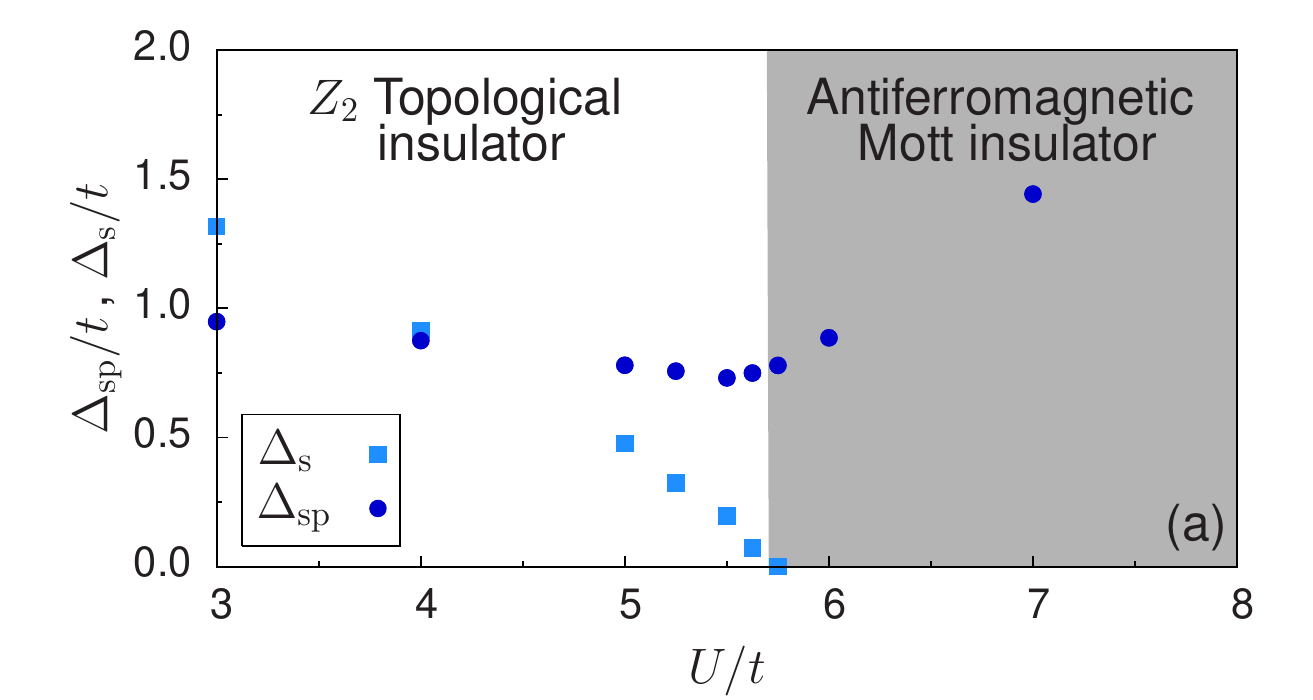}
\includegraphics[width=0.45\textwidth]{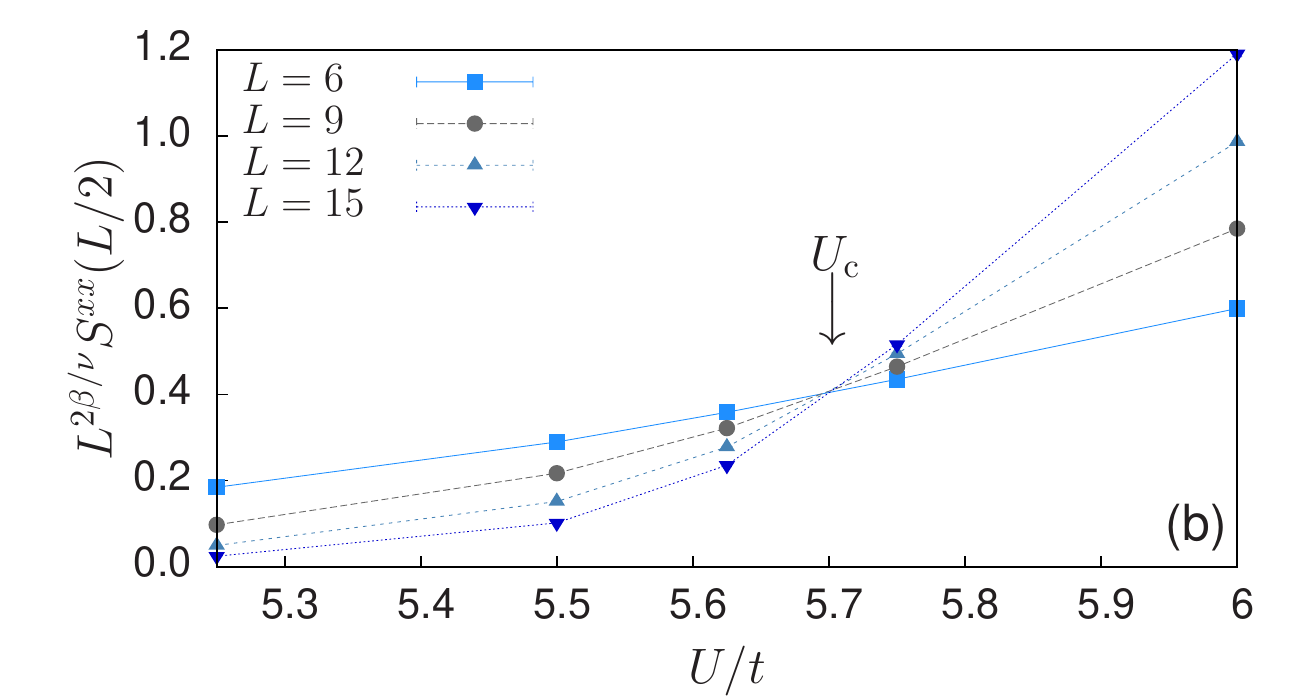}
\caption{\label{fig:magnetism} (Color online) 
  (a) Spin gap $\Delta_\text{s}(\bm{q}=0)$ and
  single-particle gap $\Delta_\text{sp}(\bm{q}=\boldsymbol{K})$ in the
  thermodynamic limit as a function of the Hubbard
  repulsion $U$, at $T=0$ ($\lambda/t=0.2$, $\alpha=0$). 
  (b) Scaling of $S^{xx}(L/2)$ using the critical exponents of the 3D XY model,
  $z=1$, $\nu=0.6717(1)$ and  $\beta=0.3486(1)$ \cite{PhysRevB.63.214503}. The
  intersection gives the critical point $U_\text{c}/t=5.70(3)$. The lattice size
  is $L\times L$. Errorbars are smaller than the symbols.}
\end{figure}

Figure~\ref{fig:freespinons} shows results for $\chi_\text{s}$ as a function
of temperature for the Kane-Mele model with two $\pi$ fluxes located at the
largest possible distance. At temperatures $\kB T\approx \Delta_\text{s}$,
$\chi_\text{s}$ is dominated by bulk effects. For $\kB T \lesssim 0.1t$, we
observe the expected Curie law.  The latter is robust with respect to
Rashba coupling, which is crucial for possible experimental realizations.

\subsection{Probing correlated topological insulators}

Figure~\ref{fig:freespinons} establishes the existence and thermodynamic signature
of degenerate spin and charge fluxons in a quantum spin Hall insulator
threaded by a pair of $\pi$ fluxes. We now consider the effect of
electron-electron interaction in the framework of the Kane-Mele-Hubbard model
(\ref{eq:KMH}). For $\lambda>0$, the phase diagram of the latter includes a
correlated quantum spin Hall insulating phase that is adiabatically connected
to that of the Kane-Mele model (\ie, $U=0$), and a Mott insulating phase with
long-range antiferromagnetic order \cite{RaHu10,Ho.Me.La.We.Mu.As.12}.
Figure~\ref{fig:magnetism}(a) shows the quantum phase transition between
these two phases as a function of $U/t$ at $\lambda/t=0.2$. At the
transition, the spin gap $\Delta_\text{s}$---as obtained from finite-size
scaling (see ref.~\onlinecite{Ho.Me.La.We.Mu.As.12} for details)---closes,
corresponding to the condensation of magnetic
excitons \cite{PhysRevLett.107.166806,Ho.Me.La.We.Mu.As.12}.
The magnetic order is of the easy-plane type,  and the transition has 3D XY
universality corresponding to the ordering of local
moments \cite{PhysRevLett.107.166806,Ho.Me.La.We.Mu.As.12}.
For $U\geq U_\text{c}$, time-reversal symmetry is spontaneously broken, and
the single-particle gap $\Delta_\text{sp}$ remains open across the
transition \cite{Ho.Me.La.We.Mu.As.12}, see Fig.~\ref{fig:magnetism}(a).

The location of the critical point can be estimated from the scaling
behavior of the real-space spin-spin correlation function
\begin{equation}\label{eq:sxx}
  S^{xx}(r) = \las S^x_\text{A}(r) S^x_\text{A}(0) \ras
\end{equation}
at the largest distance $r=L/2$. Here we consider correlations between spins
on the A sublattice, but results are the same for the B sublattice. The limit
$\lim_{L\to\infty}S^{xx}(L/2)$ is identical to $m^2$, with $m$ being the
magnetization per site. This critical value can be obtained  by considering
the 3D XY scaling behavior at the transition. Following
ref.~\onlinecite{Ho.Me.La.We.Mu.As.12}, we plot $L^{2\beta/\nu}S^{xx}(L/2)$
as a function of $U$ for different system sizes using the critical exponents
$z=1$, $\nu=0.6717(1)$ and $\beta=0.3486(1)$ \cite{PhysRevB.63.214503}.
Figure~\ref{fig:magnetism}(b) reveals the expected intersection of curves at
the critical point, and gives $U_\text{c}/t=5.70(3)$. 

\begin{figure}[t]
\includegraphics[width=0.45\textwidth]{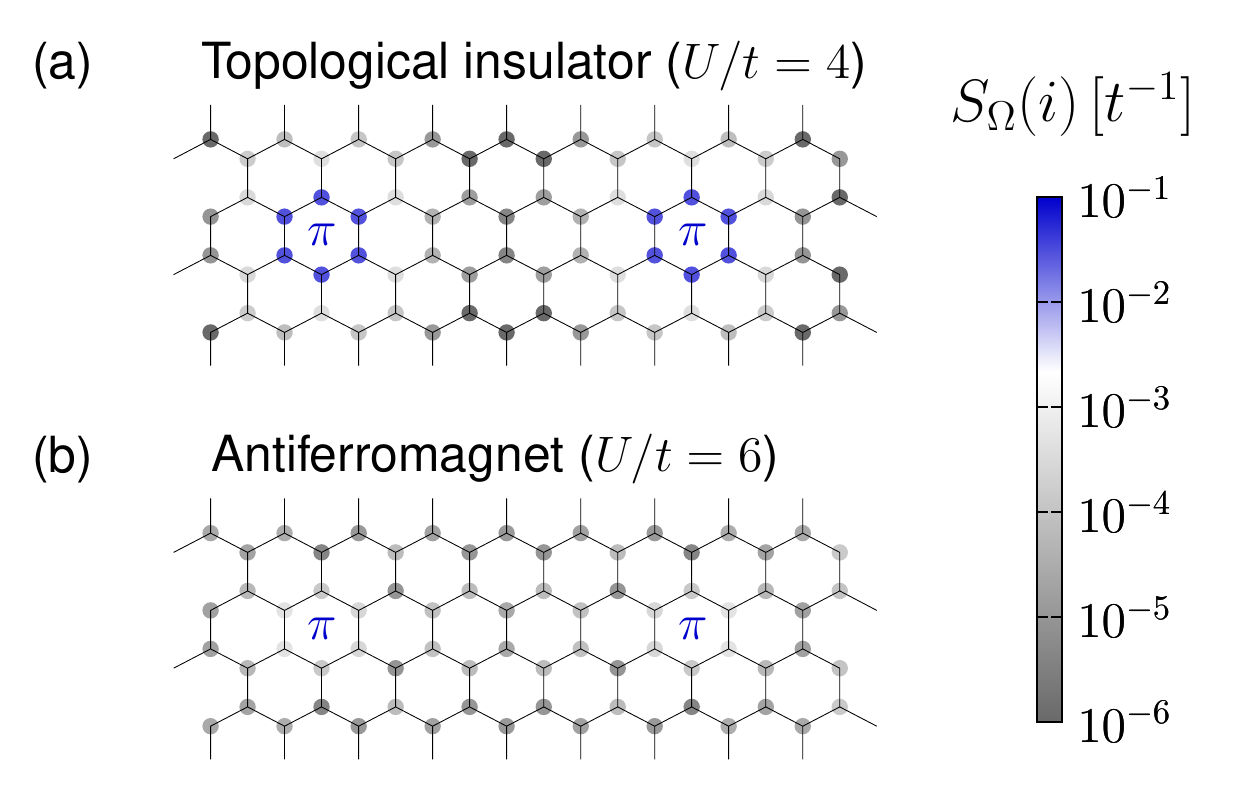}
\includegraphics[width=0.45\textwidth]{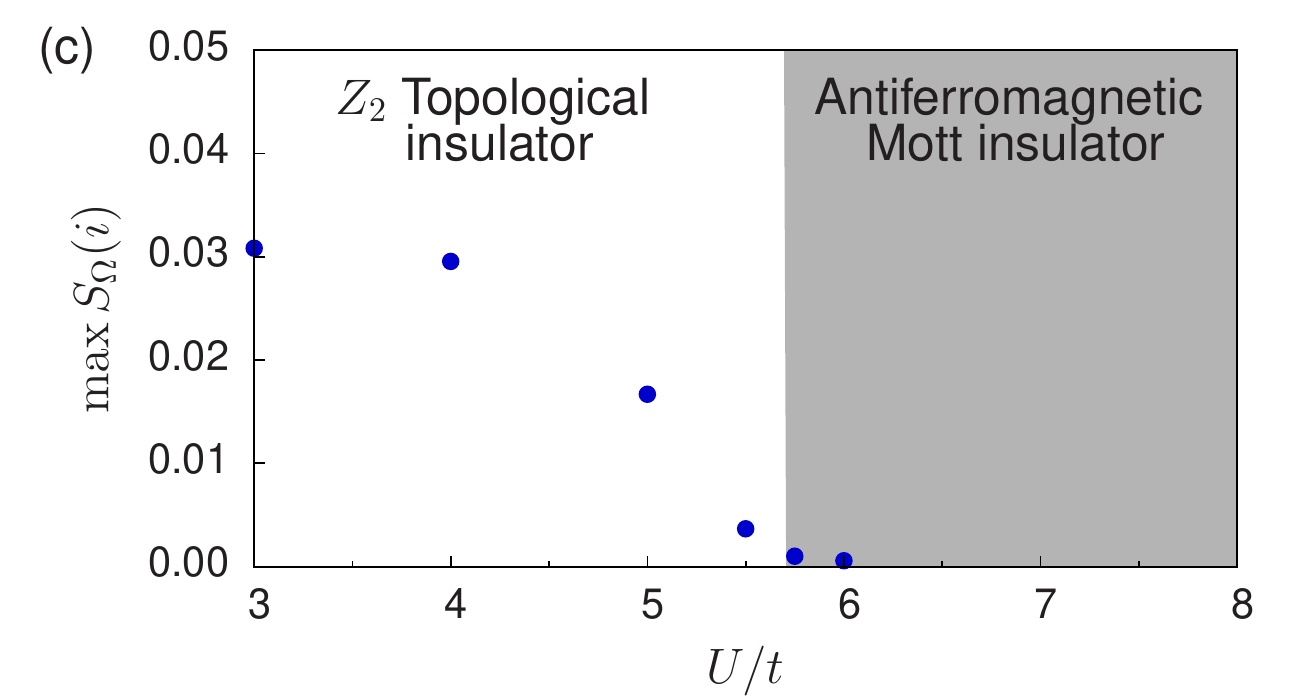}
\caption{\label{fig:localspinons}
(Color online)
Integrated dynamical spin structure factor $S_\Omega(i)$  
at $T=0$ on a $9\times9$ lattice.
(a) Localized spin fluxons created in the topological
insulator phase at $U/t=4$. 
(b) Absence of spin fluxons in the magnetic phase at
$U/t=6$.
(c) Maximum of  $S_\Omega(i)$,  as a function of $U/t$. Here
$\lambda/t=0.2$, $\alpha=0$, and $\Omega/t=0.2$. 
}
\end{figure}

\begin{figure*}[t]
\includegraphics[width=\textwidth]{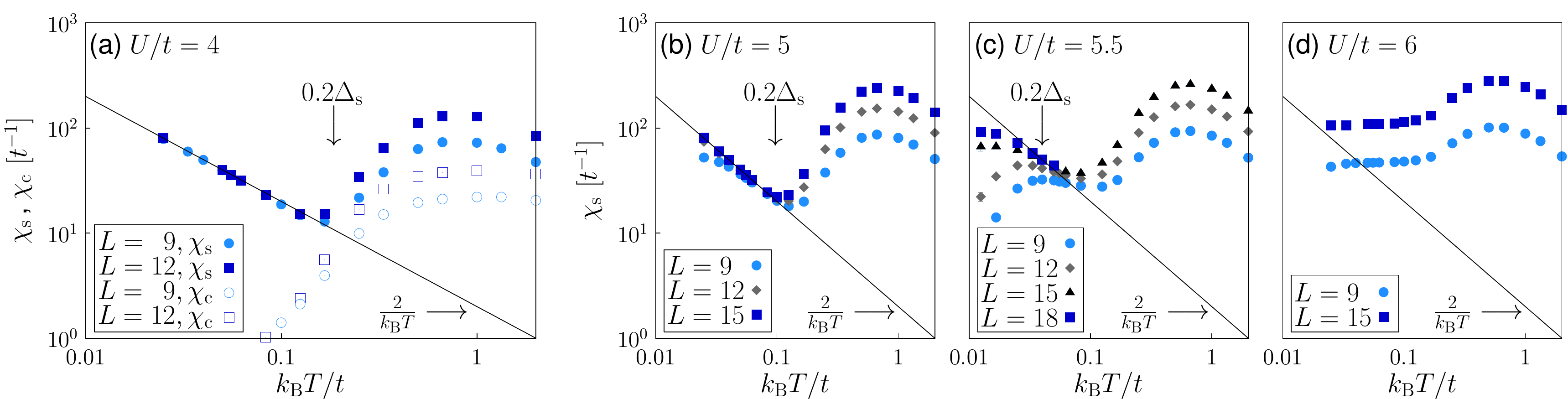}
\caption{\label{fig:twospins}
  (Color online)
  (a) Spin ($\chi_\text{s}$) and charge ($\chi_\text{c}$)
  susceptibilities of the Kane-Mele-Hubbard model
  ($\lambda/t=0.2$, $\alpha=0$) at $U/t=4$. We consider $L\times L$ lattices
  with one pair of $\pi$ fluxes placed at the maximal distance.
  At low temperatures, the spin 
  susceptibility reveals a Curie law $\chi_\text{s}=\frac{2}{\kB T}$, whereas the
  charge susceptibility is suppressed by the charge gap.
  (b)--(c), Spin susceptibility as a function of temperature for different
  values of $U/t$ ($\lambda/t=0.2$, $\alpha=0$). (a)--(c) show that
  with increasing $U/t$, the range of the interaction between spin fluxons
  increases, leading to deviations from the Curie law $\chi_\text{s}=\frac{2}{\kB T}$
  at low temperatures. 
  (d) For $U>U_\text{c}=5.70(3)t$, $\chi_\text{s}$ reflects the
  presence of long-range magnetic order in the bulk. Errorbars  are smaller
  than the symbol size. The arrows indicate the energy scale associated with the spin gap.
 }
\end{figure*}

The well-understood magnetic transition of the model~(\ref{eq:KMH}) provides
a test case for the use of $\pi$ fluxes to probe a correlated quantum spin
Hall state, as well as to track the interaction-driven transition to a
topologically trivial state. We solve
the interacting model with two $\pi$ fluxes using exact quantum Monte Carlo
simulations. Spin fluxons can be detected by calculating the lattice-site
resolved dynamical spin structure factor at $T=0$, defined as 
\begin{equation}\label{eq:dynamicalspin}
  S(i,\omega)  = \pi\sum_{n}  | \langle n |  \hat{c}^{\dagger}_{i} \sigma^{z}
  \hat{c}^\nag_{i}  | 0 \rangle |^2 \delta \left(  E_n - E_0 - \omega \right)
  \,.
\end{equation}
Here $H_\text{KMH} | n \rangle = E_n | n \rangle $, and $| 0 \rangle $ denotes
the ground state.  $S(i,\omega)$ corresponds to the spectrum of spin
excitations at lattice site $i$. A real-space map of the spin fluxon
states $\ket{\UP},\ket{\DO}$ is obtained by integrating $S(i,\omega)$ up to
an energy scale $\Omega/t=0.2$ well within the charge gap
$\Delta_\text{c}\approx 2\Delta_\text{sp}$, giving $S_{\Omega}(i)
= \int_{0}^{\Omega} \rmd \omega S(i,\omega)$. For $U/t=4$,
corresponding to the quantum spin Hall phase [see Fig.~\ref{fig:magnetism}(a)], 
we see in Fig.~\ref{fig:localspinons}(a) very sharply defined spin fluxons
localized at the two flux-threaded hexagons. The value of $S_{\Omega}(i)$
is about three orders of magnitude smaller at lattice sites which are further
away from a flux, so that the spin fluxons can easily be detected numerically. 
In Fig.~\ref{fig:localspinons}(b), we show results for the magnetic
insulating phase at $U/t=6$. As expected for this topologically trivial
state, no well-defined spin fluxons exists.

The dependence of $S_{\Omega}(i)$ on $U/t$ across the magnetic quantum phase transition
is shown in Fig.~\ref{fig:localspinons}{(c). A clear signal is found deep in
the topological insulator phase, whereas a strong drop is observed on
approaching the critical point at $U_\text{c}/t=5.70(3)$. Hence, the spin fluxon
signal can be used in quantum Monte Carlo simulations to distinguish
topological and nontopological phases.

As for the noninteracting case (Fig.~\ref{fig:freespinons}), the spin fluxons created
by the $\pi$ fluxes give rise to a characteristic Curie law in the spin
susceptibility. Figure~\ref{fig:twospins}(a) shows quantum Monte Carlo results
for the spin and charge susceptibilities defined in Eq.~(\ref{eq:susc})
in the topological phase ($U/t=4$).
We again consider two $\pi$ fluxes at the maximal separation. At low
temperatures, $\kB T\ll \Delta_\text{s}$, $\chi_\text{s}$ is well described by $\chi_\text{s}=\frac{2}{\kB
T}$, or $\frac{1}{\kB T}$ per $\pi$ flux. The additional factor of 2 compared to the
case $U=0$ comes from the splitting of spin and charge states which only
leaves the Kramers doublet $\{\ket{\UP},\ket{\DO}\}$ at low energies (see Appendix~\ref{app:twospins}).
The Curie law holds down to the lowest
temperatures considered in Fig.~\ref{fig:twospins}(a). Finally, the charge
susceptibility is strongly suppressed at low temperatures, and reveals the
absence of low-energy charge fluxons as a result of the Hubbard repulsion.

\subsection{Interaction between spin fluxons}

So far, we have exploited the thermodynamic and spectral signatures of
independent spin fluxon excitations (\ie, free spins). On periodic lattices,
spin fluxons can only be created in pairs, and it is therefore interesting to
consider their mutual interaction. Such interactions will play a key role in
Sec.~\ref{sec:spinmodels}, where we consider quantum spin Hall insulators
with multiple $\pi$ fluxes to create and simulate systems of interacting spins.

Interaction effects due to a coupling between two spin fluxons in a lattice with
one pair of $\pi$ fluxes become visible for larger $U/t$, \ie, closer to the magnetic
transition. Figures~\ref{fig:twospins}(b) and (c) show a deviation from the
Curie law below a temperature scale determined by the interaction between
spin fluxons. In the model~(\ref{eq:KMH}), this interaction is mediated by the exchange of
collective spin excitations (magnetic excitons), which are the lowest lying
excitations of the correlated topological insulator phase, and evolve into
the gapless Goldstone mode of the magnetic state. Since magnetic order is of
the easy-plane type, the dominant contribution of the resulting interaction
is expected to have the general form
\begin{align}\label{eq:Sint}
  S_\text{int} =  - g^2  &\sum_{ \bm{r}\neq \bm{r'} }  
  \iint_{0}^{\beta}  \rmd \tau \rmd \tau' \\\nonumber
  &\times
  \big[
  {\cal S}^+_{ \bm{r} }(\tau) D( \bm{r} - \bm{r}', \tau - \tau' ) 
  {\cal S}^-_{ \bm{r'} }(\tau')  + \text{H.c.}\big]\,,
 \end{align}
where ${\cal S}^\pm_{ \bm{r} }(\tau)$ are spin-flip operators acting on a
spin fluxon at position $\bm{r}$ at time $\tau$, $D(\bm{r},\tau)$ is the Fourier
transform of the exciton propagator $D( \bm{q}, \rmi\Omega_m) $  ($\bm{q}$:
momentum, $\Omega_m=2n\pi/\beta$: bosonic Matsubara
frequency), and $g$ is a coupling constant. At long wavelengths, the
dispersion relation of the collective spin mode can
be written as $\omega(\bm{q}) = \sqrt{  v^2 |\bm{q}-\bm{Q}|^2 + \Delta_s^2}$,  where
$v$ is the spin velocity,  $\Delta_s$ is the spin gap, and $\bm{Q}$ is the
magnetic ordering wavevector. The minimal exciton energy is given by
$\om(\bm{q}=\bm{Q})=\Delta_\text{s}$. Fourier transformation of the propagator
(see Appendix~\ref{app:fourier}) gives in the limit of low energies and long wavelengths
\begin{equation}\label{eq:J}
  D(\bm{r},\tau) \sim \exp(\rmi \bm{Q}\cdot \bm{r}) \exp(- \Delta_\text{s}\tau)
  \exp\left(-\frac{|\bm{r}|^2 \Delta_\text{s}}{2v^2\tau}\right)\,.
\end{equation}
The first term determines the sign of the interaction. The decay at large
imaginary time $\tau$ is governed by the spin gap $\Delta_\text{s}$. The fast
decay as a function of $|\bm{r}|$ underlies the clear Curie law seen, \eg,
in Fig.~\ref{fig:twospins}. The interaction range and strength can be tuned
via the spin gap and hence [cf. Fig.~\ref{fig:magnetism}(a)] by varying $U/t$.

From Eq.~(\ref{eq:J}) we expect the {\it interaction range} to increase
with increasing $U/t$ due to the decrease of $\Delta_\text{s}$, see
Fig.~\ref{fig:magnetism}(a). Indeed, Figs.~\ref{fig:twospins}(b) and (c)
reveal an enhanced effect of the spin fluxon separation at low temperatures with
increasing $U/t$. In particular, for $U/t=5.5$ (close to the magnetic transition),
Fig.~\ref{fig:twospins}(c) shows a Curie law corresponding to two
free spin fluxons only for the largest system sizes ($L=18$). As $U\to
U_\text{c}$, the interaction range diverges, and free spin fluxons can no longer
exist. For $U>U_\text{c}$, time-reversal invariance
is broken and $\pi$ fluxes do not create spin fluxons. Instead, the
spin susceptibility in Fig.~\ref{fig:twospins}(d) is that of an
antiferromagnet; the finite value of $\chi_\text{s}/L^2$ at $T=0$ reflects
the density of spin wave excitations.  

\begin{figure}[t]
  \includegraphics[width=0.45\textwidth]{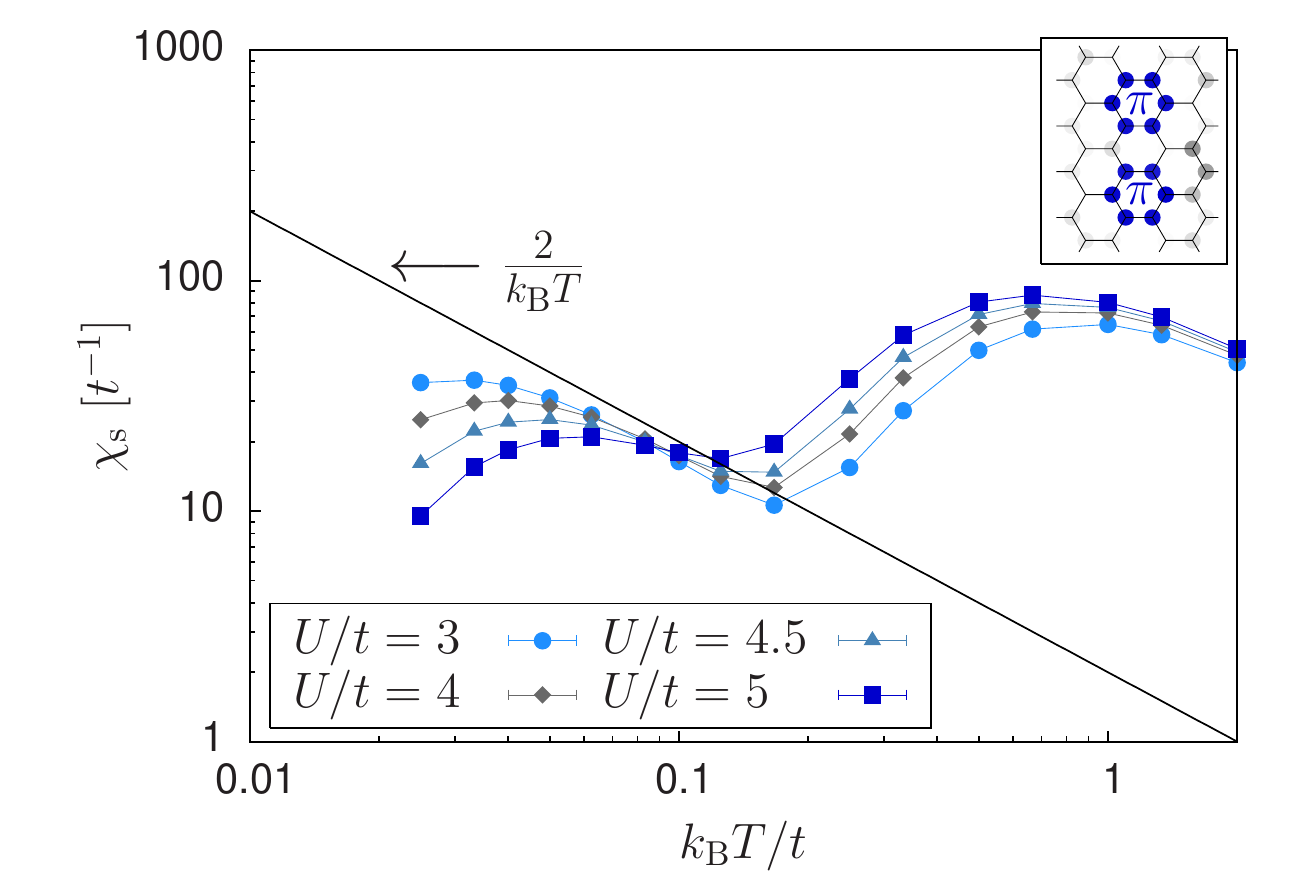}
  \caption{\label{fig:twospinsU}
    (Color online)
    Spin susceptibility as a function of temperature for {\it two} $\pi$
    fluxes arranged as shown in the inset ($\lambda/t=0.2$, $\alpha=0$,
    $9\times9$ lattice).
    With increasing $U/t$, the strength of the interaction between spin
    fluxons increases, as revealed by the shift of the
    temperature below which deviations from a $\frac{2}{\kB T}$ Curie law occur.
    Statistical errors are  smaller than the symbol size. Inset:
    $S_\Omega(i)$ for $U/t=4$, using the same color coding as in Fig.~\ref{fig:localspinons}(a).
 }
\end{figure}

To illustrate the dependence of the {\it interaction strength} on
$\Delta_\text{s}$, we consider two fluxes at a fixed, small distance as illustrated in the
inset of Fig.~\ref{fig:twospinsU}. We show the spin susceptibility for different
values of $U/t$ in Fig.~\ref{fig:twospinsU}. For $U/t=3$, a Curie
law $\chi_\text{s}\approx \frac{2}{\kB T}$ may be inferred at temperatures $\kB T\approx 0.1t$.
Increasing $U/t$, the interaction between the spin fluxons becomes too large to observe
free spin fluxons below the temperature range set by the bulk spin gap $\Delta_\text{s}$. The
downturn of $\chi_\text{s}$ occurs at higher and higher temperatures with
increasing $U/t$, and reflects a tunable, correlation-induced energy scale
for the interaction between spin fluxons that is  absent in Fig.~\ref{fig:freespinons}.

\section{$\pi$-flux quantum spin models}\label{sec:spinmodels}

The possibility of inserting $\pi$ fluxes to create localized spin fluxons
with a tunable interaction mediated by magnetic excitons provides a toolbox
to engineer interacting spin models in correlated topological insulators.
The computational effort for quantum Monte Carlo simulations does not depend
on the number of $\pi$ fluxes, and the latter can be arranged in almost arbitrary
geometries on the honeycomb lattice.

\begin{figure}[t]
  \includegraphics[width=0.45\textwidth]{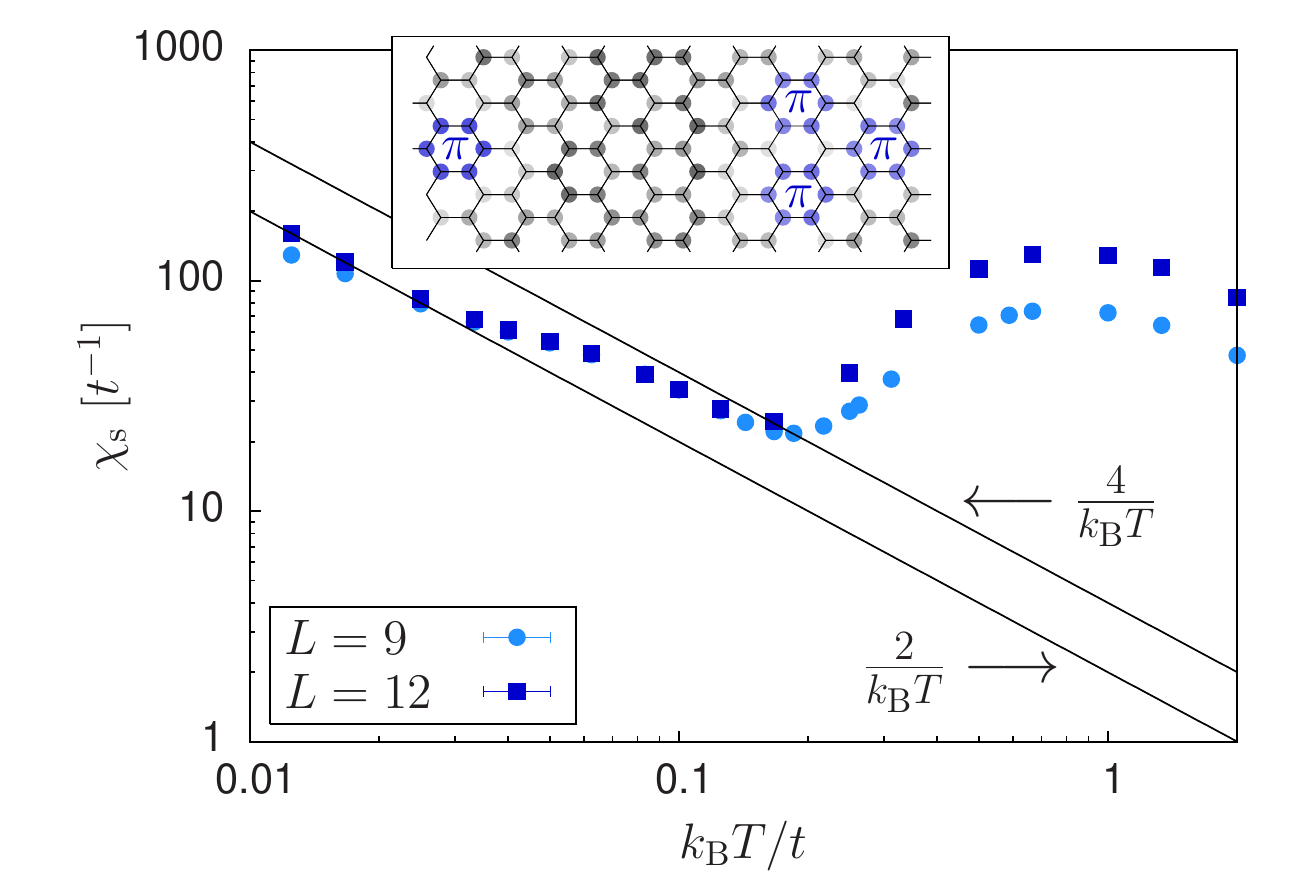}
  \caption{\label{fig:fourspins}
    (Color online)
    Spin susceptibility as a function of temperature for {\it four} $\pi$
    fluxes arranged as shown in the inset ($\lambda/t=0.2$, $\alpha=0$,
    $U/t=4$) on $L\times L$ lattices. The data reveal a
    Curie law $\frac{4}{\kB T}$ at intermediate temperatures, and $\frac{2}{\kB T}$ at low
    temperatures. Statistical errors are   smaller than the symbol size.
    Inset: $S_\Omega(i)$ for $L=15$, using the same color coding as in Fig.~\ref{fig:localspinons}(a).
 }
\end{figure}

\subsection{Three-spin system}

\begin{figure}[t]    
  \includegraphics[width=0.5\textwidth,clip]{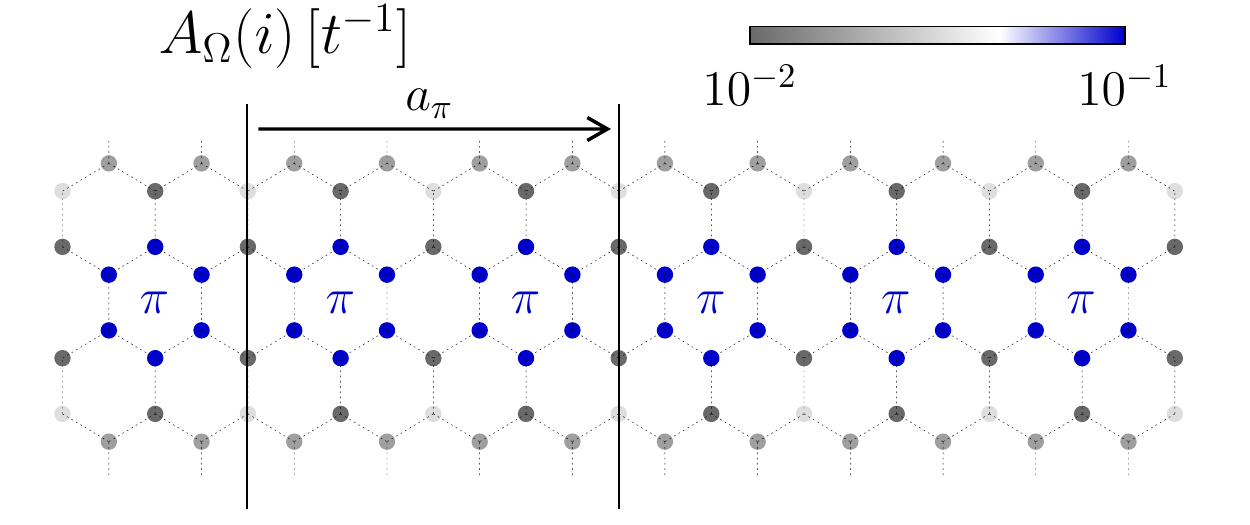}
  \caption{\label{fig:fluxchain}
    (Color online) 
    Integrated local density of states $A_\Omega(i)$
    ($\Omega=0.2t$, see text) at $T=0$ for the
    Kane-Mele model ($\lambda/t=0.2$, $\alpha=0$) with a periodic chain of $\pi$ fluxes. 
    We show a part of the $72\times 12$ lattice used, and the size of the magnetic unit
    cell containing two $\pi$ fluxes. The latter has width $a_{\pi} = 4 a$, where
    $a \equiv 1$  corresponds to the norm of the lattice vectors of
    the underlying honeycomb lattice.
 }
\end{figure}

As a first extension of the two-spin cases considered so far, 
we consider four spin fluxons emerging from two pairs
of $\pi$ fluxes. The fluxes are arranged so that three spin fluxons form a ring,
and the fourth spin fluxon is located at the largest distance from the center of
the ring. For large enough lattices, the separated spin fluxon will not couple to
the other three, and the physical problem is similar to experiments on
coupled quantum dots \cite{Gaudreau.2012} or flux qubits \cite{0295-5075-76-3-533}
in the context of quantum computation. The three spin fluxons experience a
transverse interaction of the form~(\ref{eq:Sint}), and behave as an effective
spin with $S^z=\pm1/2$ at low temperatures (see Appendix~\ref{app:fourspins}).
The spin susceptibility for $U/t=4$ shown in Fig.~\ref{fig:fourspins} reveals
that, at low temperatures, the two independent spins indeed give rise to the
expected Curie law  $\chi_\text{s}=\frac{2}{\kB T}$. At higher temperatures $\kB T
\approx 0.1t$, we find $\chi_\text{s}=\frac{4}{\kB T}$,  corresponding to four
independent spin fluxons. 
In the regime where $\chi_\text{s}=\frac{2}{\kB T}$, the sign of the
interaction between the spin fluxons determines the ground state degeneracy of the three-spin cluster. A
net ferromagnetic interaction results in a spin-$1/2$ doublet, whereas an
antiferromagnetic coupling gives rise to a four-fold degenerate, chiral
ground state (see Appendix~\ref{app:fourspins}). In principle, the sign 
of the exchange coupling for the case of Fig.~\ref{fig:fourspins} can be
determined from entropy measurements. Since $\bm{Q}=0$ for the model~(\ref{eq:KMH}),
Eq.~(\ref{eq:J}) suggests that the interaction is ferromagnetic.

\subsection{Simulation of one-dimensional fluxon chains}

Whereas the study of systems with a small number of spins is relevant for
applications such as quantum computing, many questions in condensed
matter physics are related to  periodic spin lattices. In this
section, we therefore consider one-dimensional chains of $\pi$ fluxes
in the honeycomb lattice with periodic boundary conditions. 

\begin{figure}[t]    
  \includegraphics[width=0.5\textwidth,clip]{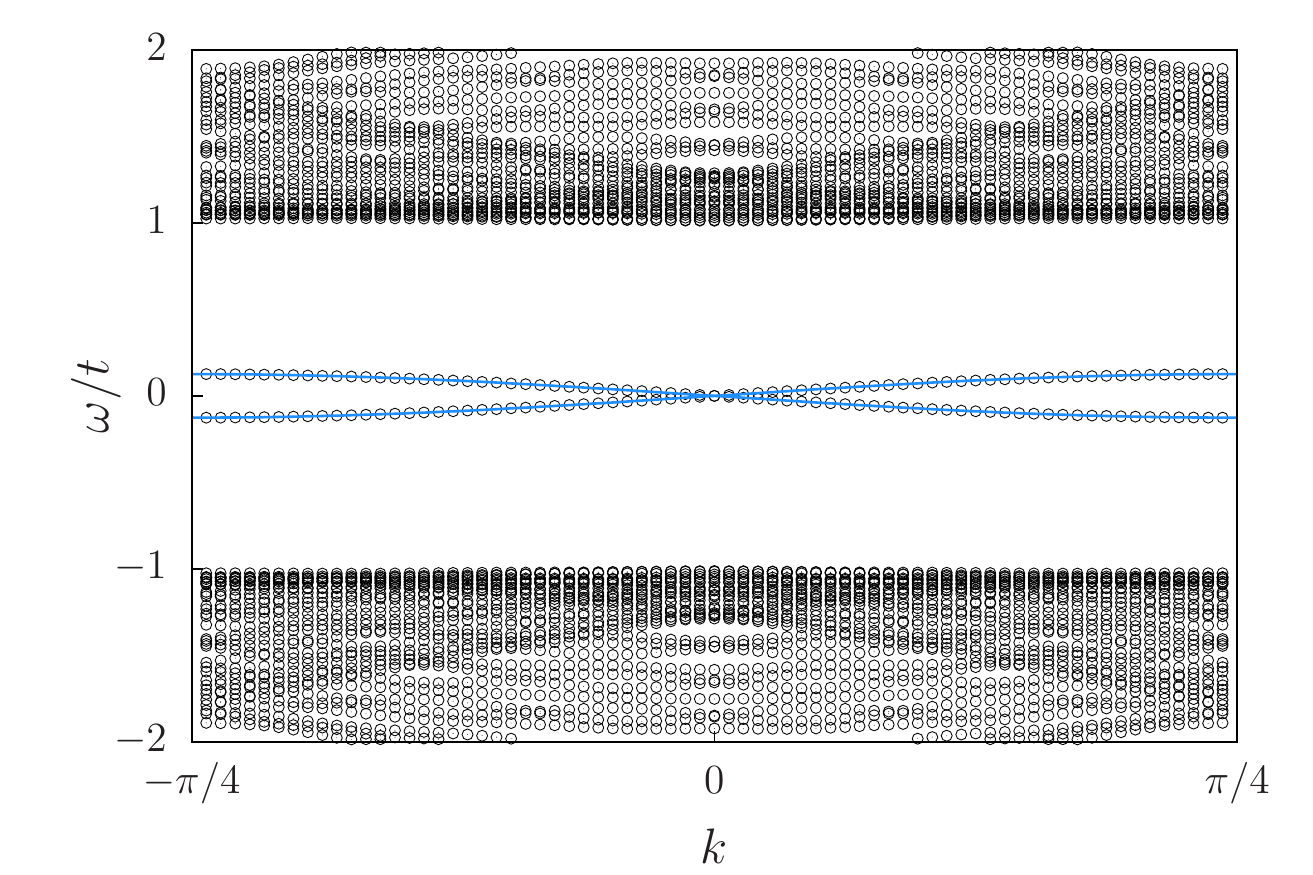}
  \caption{\label{fig:1dspectrum}
    (Color online)
    Spectrum of eigenvalues of the Kane-Mele model with a 
    periodic chain of $\pi$ fluxes (cf. Fig.~\ref{fig:fluxchain}).  Here $\lambda/t=0.2$,
    $\alpha=0$, and the honeycomb lattice has dimensions $72\times12$. Points
    correspond to eigenvalues, and lines to the band
    structure $\epsilon(k)=\pm2\tilde{t} \sin(2k a)$ with
    $\tilde{t}\approx0.126t$ and $ a \equiv 1$.
   }
\end{figure}

We begin with the noninteracting Kane-Mele model with a periodic flux chain.
The fluxon excitations are
visible in Fig.~\ref{fig:fluxchain} which shows the integrated local density
of states $A_\Omega(i)  = \int_{0}^{\Omega} d\omega A(i,\omega)$; the
single-particle spectral function $A(i,\omega)$ is defined as usual
in terms of the single-particle Green function, $A(i,\omega) = -\text{Im}\,
G(i,\omega)$. Whereas the fluxons are  well localized in the direction normal
to the chain, the overlap of neighboring
fluxons in the chain gives rise to a tight-binding band inside the
topological band gap which can be seen in the spectrum shown in
Fig.~\ref{fig:1dspectrum}. The specific form of the band structure 
can be attributed to the fact that the smallest unit cell for the fluxon
chain contains two flux-threaded hexagons (and is four hexagons wide), see
Fig.~\ref{fig:fluxchain}. Exploiting that the four possible fluxon states per
hexagon, $\{\ket{\UP},\ket{\DO},\ket{+},\ket{-}\}$, can formally be written
in terms of the fermion Fock states $\{\ket{\UP},\ket{\DO},\ket{0},\ket{\UP\DO}\}$,
and assuming nearest-neighbor hopping, a suitable Hamiltonian is given by
\begin{equation}
  H = -\tilde{t}\sum_{i\sigma} (\phi^\dag_{i\sigma} \psi^\nag_{i\sigma} - 
  \psi^\dag_{i\sigma} \phi^\nag_{i+a_{\pi}\sigma} + \text{H.c.})\,,
\end{equation}
where $\phi$, $\psi$ refer to the two flux-threaded hexagons in the unit cell,
and $i$ numbers the unit cells. The resulting band dispersion
$\epsilon(k)=\pm 2\tilde{t} \sin(2k a)$ matches the low-energy bands in
the spectrum (Fig.~\ref{fig:1dspectrum}).  The form of the 
effective low energy Hamiltonian and especially the gapless  nature of the
spectrum  stems from  the fact that the unit cell is a gauge choice; a
translation  by half a lattice vector, $a_{\pi}/2$, corresponds to a gauge
transformation.   This symmetry allows the intercell and intracell
hopping integrals to differ  only by a phase $e^{i \theta}$.  Imposing time-reversal
symmetry pins the phase factor to $\theta= 0$ and $\theta=\pi$,  thus leading to
the dispersion relations $ \pm 2\tilde{t} \cos\left[ ( k + \theta/a_{\pi})
  a_\pi/2 \right]$. The choice $\theta= \pi$ produces the above mentioned
dispersion relation, and the choice $\theta = 0$ merely corresponds to
translating the  reciprocal lattice by half  a reciprocal lattice unit vector.

In contrast to the helical edge states of a quantum spin Hall insulator,
each of the two fluxon bands is spin degenerate. As a result, and because
the system is half filled, we expect a Mott transition of charge fluxons for
any nonzero electron-electron repulsion.
Figure~\ref{fig:1dchi} shows the spin and charge
susceptibilities of the Kane-Mele-Hubbard model on $L\times12$ lattices
with $L/2$ $\pi$ fluxes and $U/t=4$. The Hubbard $U$ causes an exponential
suppression of the charge susceptibility at low temperatures [see
Fig.~\ref{fig:1dchi}(b) and inset], whereas low-energy spin fluxon excitations
remain [Fig.~\ref{fig:1dchi}(a)]. Hence, similar to the one-dimensional
Hubbard model, the fluxon chain undergoes a Mott transition to a state with a
nonzero charge gap but gapless spin excitations.

\begin{figure}[t]
\includegraphics[width=0.45\textwidth]{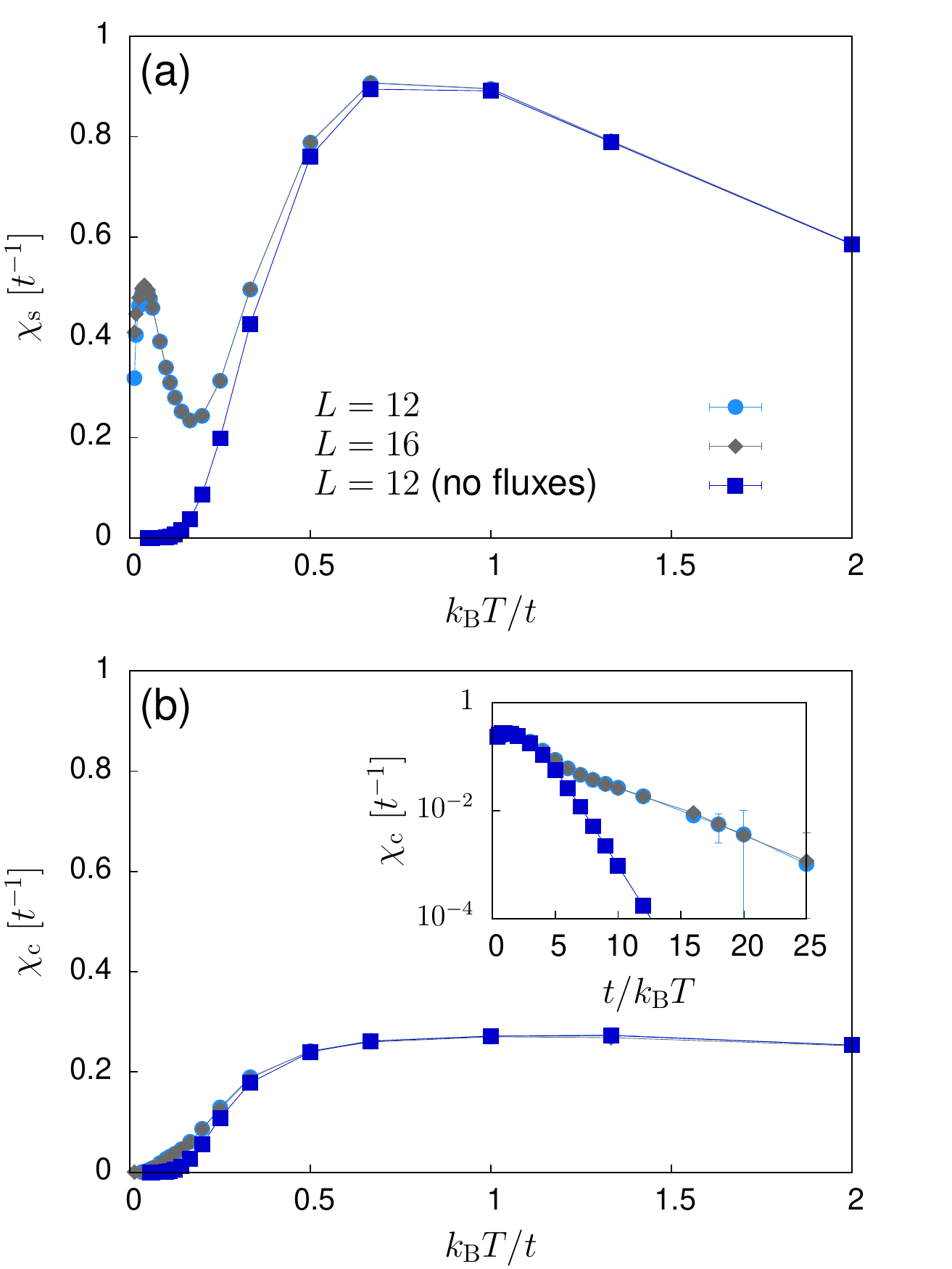}
\caption{\label{fig:1dchi}
  (Color online)
  (a) Spin and (b) charge  susceptibility of the Kane-Mele-Hubbard model
  ($\lambda/t=0.2$, $\alpha=0$) at $U/t=4$. We consider $L\times 12$ lattices
  with $L/2$ $\pi$ fluxes arranged in a periodic one-dimensional chain.
  The inset in (b) shows the charge susceptibility as a function of
  inverse temperature on a logarithmic scale. The key in (a) applies to all
  panels.
 }
\end{figure}

In the Mott phase of the fluxon chain, the low-energy physics is expected to
be described by spin fluctuations, and hence by an effective spin model with
spins corresponding to Kramers doublets of localized spin fluxons. Because the
interaction range depends exponentially on the spin gap, we
expect nearest-neighbor interactions $J^{xy}$, $J^{zz}$ between spin fluxons to
dominate, except for the close vicinity of the magnetic transition.  As argued
before, the magnetic exciton is of predominantly easy-plane type, 
and we therefore expect anisotropic interactions, $|J^{xy}|\gg |J^{zz}|$.
The minimal model for the spin chain is the one-dimensional XXZ Hamiltonian,
\begin{equation}
  H = J^{zz} \sum_i S^z_i S^z_{i+1} + J^{xy} \sum_i (S^+_i S^-_{i+1} + S^-_i S^+_{i+1})\,.
\end{equation}
Using the ALPS 1.3 implementation \cite{ALPS_I} we can simulate this
model in the stochastic series expansion representation to calculate the spin
susceptibility as a function of temperature. There is one free parameter,
$J^{xy}/J^{zz}$, which is varied to obtain the best fit to the
low-temperature susceptibility (at high temperatures, bulk  states of the
topological insulator begin to play a role) of the Kane-Mele-Hubbard model.
For example, considering six
spins, a rather good match between the spin fluxon data and the XXZ model is
obtained for $J^{zz}/|J^{xy}| = -0.1$ (the sign of $J^{xy}$ is irrelevant), see
Fig.~\ref{fig:spinchain}. Importantly, taking the same parameters, and
simulating ten spins with both spin fluxons and the XXZ model, equally good
agreement is found in Fig.~\ref{fig:spinchain}. These results demonstrate
that the spin fluxons form a one-dimensional spin system with well-defined
interactions, and that a quantum spin Hall insulator with $\pi$
fluxes can indeed be used as a quantum simulator. 

\begin{figure}[t]    
  \includegraphics[width=0.45\textwidth]{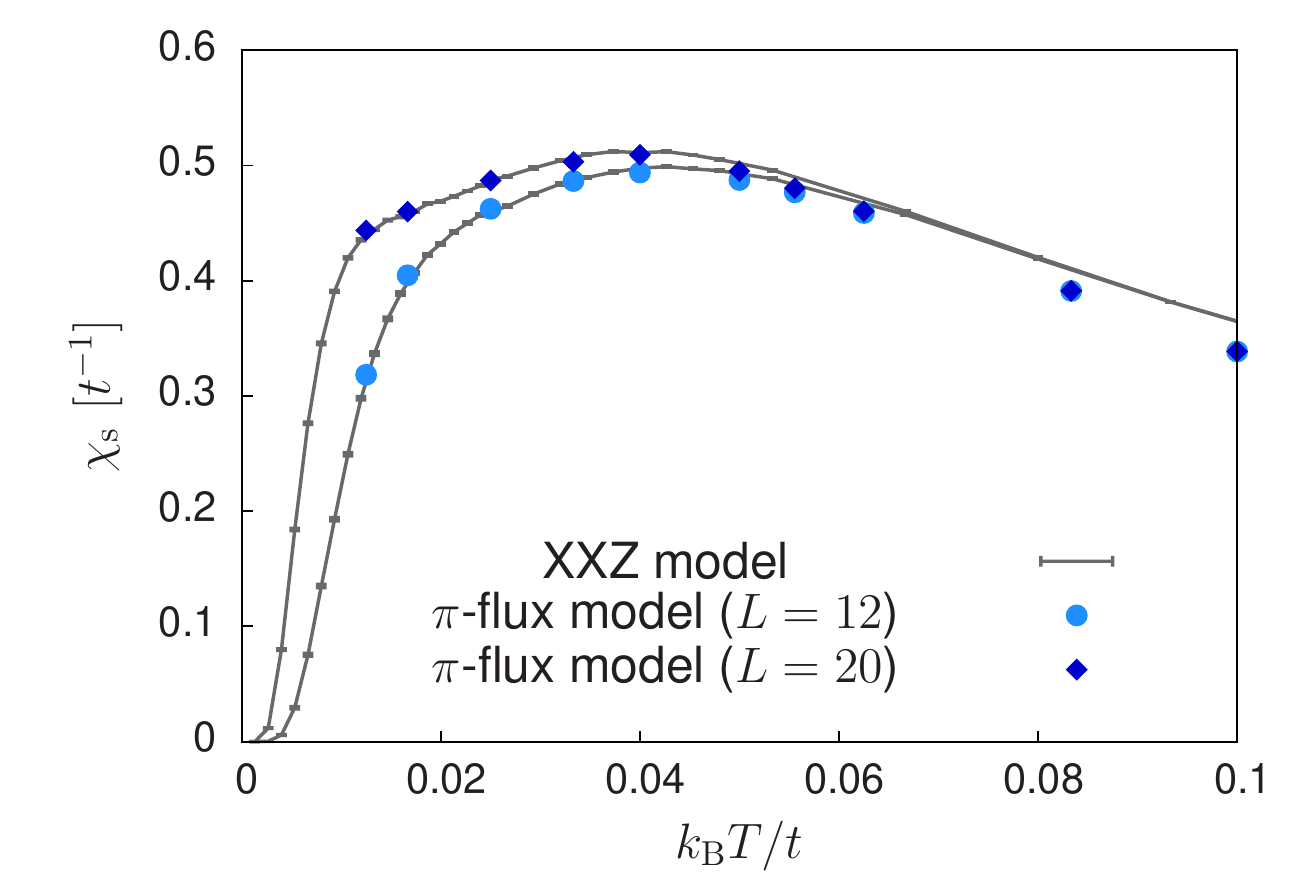}
  \caption{\label{fig:spinchain}
    (Color online)
    Spin susceptibility as a function of temperature. Symbols correspond to quantum Monte Carlo results for 
    the Kane-Mele-Hubbard model ($\lambda/t=0.2$, $\alpha=0$, $U/t=4$)
    on $L\times 12$ lattices with $L/2$ $\pi$ fluxes arranged in a
    periodic one-dimensional chain. Lines are quantum Monte Carlo results
    for the one-dimensional XXZ model with $J^{zz}/|J^{xy}|=-0.1$ and $L/2$
    lattice sites (spins).
 }
\end{figure}

\section{Conclusions}\label{sec:conclusions}

In this work, we have presented quantum Monte Carlo results for a correlated
quantum spin Hall insulator with topological defects in the form of $\pi$
fluxes. Such fluxes represent a universal probe for the topological index
that can be used in the presence of electronic correlations, and does not
rely on spin conservation or an adiabatic connection to a noninteracting
topological insulator. Our results demonstrate that $\pi$ fluxes can
be combined with exact numerical simulations, and lead to clear signatures
of nontrivial topological properties in spectral and thermodynamic properties.
As a concrete example, we have studied the magnetic quantum phase transition
of the Kane-Mele-Hubbard model at which time-reversal symmetry is spontaneously broken.
In principle,  $\pi$ fluxes can also be used in connection with fractional quantum spin
Hall states.

More generally, $\pi$ fluxes in correlated topological insulators allow one
to construct and simulate quantum spin models, and hence lead to a novel
class of quantum simulators. This finding is not restricted to the
Kane-Mele-Hubbard model considered here. In particular, magnetism driven by
electronic correlations---the origin of the interaction between spin fluxons---is a
common phenomenon. The physics described here relies on
the coexistence of magnetic correlations and time-reversal symmetry, and
cannot be captured by static mean-field descriptions. The spin models share the topological
protection of their host against, for example, disorder. In general, they are
characterized by a dynamical, time-dependent interaction  reminiscent of
spin-boson problems.  The detailed form and sign of the interaction, whose
strength and range can be tuned via the electronic interactions, depends on
the electronic Hamiltonian and the lattice geometry of the underlying
topological insulator.  Because of the spin-orbit interaction, the spin symmetry is
$U(1)$, and---similar to cold atom realizations of the quantum Ising model
\cite{PhysRevLett.103.265302}---the spin-spin interaction is generically
anisotropic. We have provided explicit evidence for the feasibility of our
idea in terms of simulations of two and four spins, as well as of one-dimensional spin chains.
With additional Rashba terms, spin models with a discrete $Z_2$ Ising
symmetry result. Although spin fluxon states are still well
defined \cite{PhysRevLett.101.086801,Qi08}, it is {\it a priori} not
clear which operators have to be measured in the numerical
simulations. Finally, the concept of fluxons originating from $\pi$ fluxes
carries over to three-dimensional topological
insulators \cite{PhysRevLett.101.086801,PhysRevB.82.041104}.

An open question of central importance is whether the use of $\pi$ fluxes
will enable us to study quantum spin systems which are currently not 
accessible to numerical methods, for example due to a sign problem in the
presence of frustrated interactions. Whereas we have provided evidence for
the possibility to simulate arrays and chains of quantum spins, and to
tune the interaction strength and range, entropy measurements
are required to determine the sign of the interactions. However, the latter
are extremely demanding to carry out using discrete-time quantum Monte Carlo
methods. A systematic effort to study spin fluxon chain and ladder geometries
is currently in progress.

Our idea may potentially also be used in experiments. A strongly correlated
topological insulator on the honeycomb lattice may be realized
with \chem{Na_2IrO_3} \cite{Irridates-Nagaosa}, or with molecular graphene
\cite{Go.Ma.Ko.Gu.Ma.12}. It has been suggested that $\pi$ fluxes can be
created in a quantum spin Hall insulator
by means of an adjacent superconductor and a magnetic field \cite{Qi08}. This
idea can be generalized to arrays of $\pi$ fluxes using Abrikosov lattices.
Alternatively, $\pi$ fluxes may be realized using SQUIDs. A potential problem
is that the diameter of the $\pi$ fluxes will not be of
the order of the lattice constant. Other exciting recent proposals which are
relevant for the realization of our idea include artificial semiconductor
honeycomb structures \cite{Singha11}, cold atoms in optical
lattices \cite{PhysRevLett.107.145301}, and cold atoms on
chips \cite{PhysRevLett.105.255302}. In solid state setups, $\pi$ fluxes can
also be created by dislocations \cite{Ran-dislocations,Zaanen12} or
wedge disclinations \cite{RueggLin12}.

{\begin{acknowledgments}%
We thank N. Cooper, T. L. Hughes, L. Molenkamp, J. Moore, J. Oostinga,
X.-L. Qi, C. Xu, and S. C. Zhang for insightful conversations, and
acknowledge support from DFG Grants No.~FOR1162 and As~120/4-3, as well
as generous computer time at the LRZ Munich and the JSC. We made use
of ALPS 1.3 \cite{ALPS_I}.
\end{acknowledgments}}

\appendix

\section{Spin susceptibility for two $\pi$ fluxes}\label{app:twospins}

A single $\pi$ flux in a topological insulator gives rise to four states,
$\ket{\UP},\ket{\DO},\ket{+},\ket{-}$. In the absence of correlations, these
states are degenerate.  At low temperatures, the spin susceptibility, defined
in Eq.~(\ref{eq:susc}), can be calculated using the Hilbert space formed
by only these states. Defining an effective Hamiltonian $H_\pi = \sum_\psi E^\psi
\ket{\psi}\bra{\psi}$ with $\psi\in\{+,-,\UP,\DO\}$ and $E^+=E^-=E^\UP=E^\DO=E^{\UP\DO}$, we obtain
\begin{align}
  \chi_\text{s} 
  &=
  \beta \left( \langle \hat{M}_z^2 \rangle - \langle \hat{M}_z \rangle^2
  \right)
  \\\nonumber
  &=  
  \frac{1}{\kB T}
  \frac{
    \sum_{\psi} \bra{\psi} \hat{M}_z^2 e^{-\beta H_\pi} \ket{\psi}
  }
  {
    \sum_{\psi}\bra{\psi} e^{-\beta H_\pi} \ket{\psi}
  }
  =
   \frac{1}{2\kB T}
   \,.
\end{align}

For $U\gg \kB T$, the spin fluxons $\ket{\UP},\ket{\DO}$ are the only low-energy
excitations, and $\chi_\text{s}$ can be calculated by restricting $\psi$ to
$\{\UP,\DO\}$. Since $E^\UP=E^\DO=E^{\UP\DO}$ due to time-reversal symmetry,
we get
\begin{equation}
  \chi_\text{s} 
  =
  \frac{1}{\kB T}
  \,.
\end{equation}

For the case of two independent $\pi$ fluxes, the above results imply
$\chi_\text{s}=\frac{1}{\kB T}$ at $U=0$, and $\chi_\text{s}=\frac{2}{\kB T}$
for $U>0$. This agrees with the numerical results shown in
Fig.~\ref{fig:freespinons} for $U=0$, and in
Figs.~\ref{fig:twospins},\ref{fig:twospinsU} for $U>0$. 

Our derivation is only valid in the absence of Rashba spin-orbit coupling
$\alpha$. However, the numerical results in Fig.~\ref{fig:freespinons} show that
the low-temperature Curie law in $\chi_\text{s}$ is the same also for $\alpha\neq0$.

\section{Spin susceptibility and ground state degeneracy for four $\pi$
  fluxes}\label{app:fourspins}

The results for the Kane-Mele-Hubbard model with four $\pi$ fluxes
shown in Fig.~\ref{fig:fourspins} reveal a $\frac{2}{\kB T}$ Curie law at low
temperatures, and a $\frac{4}{\kB T}$ Curie law at higher temperatures. This finding
can be understood as corresponding to either two or four noninteracting
spins. The latter case corresponds to the spatially separate spin fluxon and an effective
spin-$1/2$ Kramers doublet (formed by the three nearby spin fluxons) in the regime where
$\chi_\text{s}\approx\frac{2}{\kB T}$, and to four noninteracting spin fluxons in the
regime where $\chi_\text{s}\approx\frac{4}{\kB T}$.

The cluster formed by the three nearby spin fluxons has the possible configurations
\begin{align}
  |M^z| = \threeh:\quad &\{\ket{\UP \UP \UP},  \ket{\DO \DO \DO}\}\,,\\\nonumber
  |M^z| = \oh:\quad
  &\{\ket{\UP \DO \DO},  \ket{\DO \UP \DO},  \ket{\DO \DO \UP},  \ket{\DO \UP \UP},  \ket{\UP \DO \UP},  \ket{\UP \UP \DO}\}\,,
\end{align}
where $M^z$ denotes the total spin in the $z$ direction. Since the
exciton-mediated interaction in the Kane-Mele-Hubbard model has the
form given in Eq.~(\ref{eq:Sint}) and hence promotes spin-flips,
the ground state can be expected to have $|M^z|=1/2$. The above mentioned
effective spin-$1/2$ then corresponds to the two possible values $M^z=\pm1/2$.

The degeneracy of the ground state depends on the sign of the interaction.
Considering $M^z=1/2$, we have the allowed states  $\ket{\DO \UP \UP}$,  $\ket{\UP
\DO \UP}$ and $\ket{\UP \UP \DO}$. The spin-flip terms which connect these
states are of the form $J (S^+_{i+1} S^-_i + S^+_{i} S^-_{i+1})$, with
periodic boundary conditions. An equivalent representation is given by the
Hamiltonian
\begin{equation}
  H = J \sum_j \left(\ket{j+1}\hspace*{-0.25em}\bra{j} + \ket{j}\hspace*{-0.25em} \bra{j+1}\right)\,,
\end{equation}
which describes the hopping of a particle (the spin down) on a three-site
ring, with $\ket{1}=\ket{\DO\UP\UP}$ etc.
The eigenstates are obtained by Fourier transformation, and have the form
\begin{equation}
  \ket{k} = \frac{1}{\sqrt{3}} \sum_{j=1}^3 e^{\rmi k j} \ket{j}\,,
  \quad k = 0, \pm \frac{2\pi}{3}\,.
\end{equation}
The eigenvalues are given by
\begin{equation}
  E(k) = 2 J \cos k\,.
\end{equation}
For $J<0$, the ground state has $k=0$ and energy $2J$. For $J>0$, the
ground state is chiral, with $k=\pm 2\pi/3$ and energy $-J$. Taking into
account the sector $M^z=-1/2$, we find a total ground state degeneracy
of {\it two} in the ferromagnetic case $J<0$, and {\it four} in the
antiferromagnetic case $J>0$. 

\section{Fourier transform of the exciton propagator}\label{app:fourier}

The exciton propagator in Eq.~(\ref{eq:Sint}) takes the form
\begin{equation}
  D(\bm{q},\tau) = \frac{e^{\tau\om(\bm{q})}}{e^{\beta\om(\bm{q})}-1}
  -\frac{e^{-\tau\om(\bm{q})}}{e^{-\beta\om(\bm{q})}-1}
\end{equation}
with the exciton energy
\begin{equation}
  \omega(\bm{q}) 
  = 
  \sqrt{  v^2 |\bm{q} - \bm{Q}|^2  + \Delta_s^2} \,.
\end{equation}
In the low-temperature limit $\beta\to\infty$, the propagator becomes
$D(\bm{q},\tau)=e^{-\tau\om(\bm{q})}$. Setting $\bm{q}'=\bm{q}-\bm{Q}$, the
Fourier transform is given by
\begin{equation}\label{eq:FT}
D(\bm{r},\tau) = e^{\rmi\bm{Q}\cdot\bm{r}}\int \rmd^2 q' e^{\rmi \bm{q}'\cdot \bm{r}} e^{-\om(\bm{q}'+\bm{Q})\tau}\,.
\end{equation}
Assuming $\Delta_\text{s}\gg v$, we can expand to obtain 
\begin{equation}
  \omega(\bm{q}'+\bm{Q})
  \approx
  \Delta_s \left[1 + \frac{v^2  |\bm{q}'|^2}{2\Delta_s^2}\right]
  \,.
\end{equation}
Taking the continuum limit, the Fourier transform involves the product of two
Gaussian integrals, and the result is given by Eq.~(\ref{eq:J}).


%

\end{document}